\documentclass[fleqn,usenatbib]{mnras}

\usepackage{mathptmx}

\usepackage{xspace}

\usepackage[T1]{fontenc}

\DeclareRobustCommand{\VAN}[3]{#2}
\let\VANthebibliography\thebibliography
\def\thebibliography{\DeclareRobustCommand{\VAN}[3]{##3}\VANthebibliography}

\usepackage{graphicx}	
\usepackage{amsmath}	
\usepackage{amssymb}
\usepackage{xcolor}	
\usepackage{subcaption}

\newcommand{\msun}{\ensuremath{{\rm M}_{\sun}}\xspace}

 \title[Helium as a signature of the double detonation]{Helium as a signature of the double detonation in Type Ia supernovae}

\author[C. E. Collins et al.]{Christine E. Collins,$^{1,2}$\thanks{E-mail: c.collins@gsi.de}
    Stuart A. Sim,$^{2}$
    Luke. J. Shingles,$^{1,2}$
    Sabrina~Gronow,$^{3,4}$
    Friedrich~K.~R\"{o}pke,$^{4,5}$
    \newauthor R\"{u}diger Pakmor,$^{6}$
    Ivo R. Seitenzahl$^{7}$
    and Markus Kromer$^{4}$
    \\
    $^{1}$GSI Helmholtzzentrum f\"{u}r Schwerionenforschung, Planckstraße 1, 64291 Darmstadt, Germany\\
    $^{2}$Astrophysics Research Center, School of Mathematics and Physics, Queen's
    University Belfast, Belfast BT7 1NN, Northern Ireland, UK\\
    $^{3}$Zentrum f\"ur Astronomie der Universit\"at Heidelberg,
    Astronomisches Rechen-Institut, M\"{o}nchhofstr. 12-14, 69120 Heidelberg, Germany\\
    $^{4}$Heidelberger Institut f\"{u}r Theoretische Studien,
    Schloss-Wolfsbrunnenweg 35, 69118 Heidelberg, Germany\\
    $^{5}$Zentrum f\"ur Astronomie der Universit\"at Heidelberg, Institut f\"ur
    theoretische Astrophysik, Philosophenweg 12, 69120 Heidelberg, Germany\\
    $^{6}$Max-Planck-Institut f\"{u}r Astrophysik, Karl-Schwarzschild-Str. 1, D-85748, Garching, Germany\\
    $^{7}$School of Science, University of New South Wales, Australian Defence Force Academy, Canberra, ACT 2600, Australia\\
}

 \date{Accepted XXX. Received YYY; in original form ZZZ}

 \pubyear{2023}

\begin{document}
\label{firstpage}
\pagerange{\pageref{firstpage}--\pageref{lastpage}}
\maketitle

\begin{abstract}
The double detonation is a widely discussed mechanism to explain Type Ia supernovae from explosions of sub-Chandrasekhar mass white dwarfs.
In this scenario, a helium detonation is ignited in a surface helium shell on a carbon/oxygen white dwarf, which leads to a secondary carbon detonation.
Explosion simulations predict high abundances of unburnt helium in the ejecta, however, radiative transfer simulations have not been able to fully address whether helium spectral features would form. 
This is because helium can not be sufficiently excited to form spectral features by thermal processes, but can be excited by collisions with non-thermal electrons, which most studies have neglected.
We carry out a full non-local thermodynamic equilibrium (non-LTE) radiative transfer simulation for
{an instance of} a double detonation explosion model, and include a non-thermal treatment of fast electrons.
We find a clear \ion{He}{I} $\lambda 10830$ feature which is strongest in the first few days after explosion and becomes weaker with time.
Initially this feature is blended with the \ion{Mg}{II} $\lambda 10927$ feature but over time separates to form a secondary feature to the blue wing of the \ion{Mg}{II} $\lambda 10927$ feature.
We compare our simulation to observations of iPTF13ebh, which showed a similar feature to the blue wing of the \ion{Mg}{II} $\lambda 10927$ feature, previously identified as \ion{C}{I}.
Our simulation shows a good match to the evolution of this feature and we identify it as high velocity \ion{He}{I} $\lambda 10830$.
This suggests that \ion{He}{I} $\lambda 10830$ could be a signature of the double detonation scenario.
\end{abstract}

\begin{keywords}
radiative transfer -- line: formation -- methods: numerical -- transients: supernovae -- white dwarfs
\end{keywords}



\section{Introduction}

Type Ia supernovae (SNe Ia) are the thermonuclear explosion of a white dwarf,
but the exact mechanism by which the white dwarf explodes is not yet
fully understood.
The double detonation is a promising scenario to explain SNe Ia for a range 
of luminosities. 
In this scenario, a helium detonation is ignited in a helium shell on a carbon-oxygen white dwarf. 
The helium detonation then ignites a secondary carbon detonation in the core.
Early double detonation models \citep[see eg.][]{taam1980a, nomoto1980a, nomoto1982a, livne1990a, woosley1994b,
hoeflich1996a, nugent1997a}
invoked relatively massive helium shells ($\sim 0.2$ \msun),
and predicted light curves and spectra that were not consistent with those of normal SNe Ia,
due to over production of iron group elements in the outer ejecta.
There has been renewed interest in the double detonation following
suggestions that considerably lower mass helium shells ($< 0.1$\msun) may be able to ignite
a secondary core detonation \citep{bildsten2007a, shen2009a, fink2010a, shen2010a},
and significantly reduce the discrepancies with observations caused by the over-production
of iron group elements at high velocities, produced in the helium
shell detonation \citep{kromer2010a, woosley2011b}.
Simulations by \citet{townsley2019a} and \citet{shen2021b} have suggested that
by considering minimal helium shell masses, 
double detonations may be able to account for normal SNe Ia, although \citet{collins2022b} show that a minimal He shell mass does not necessarily lead to good agreement with normal SNe~Ia.
Additionally, \citet{glasner2018a} have shown for the first time
conditions leading to the ignition of a helium detonation in an accreted helium envelope
on top of a carbon-oxygen white dwarf.

A number of objects have recently been observed that have specifically
been suggested to be produced by helium shell detonations,
for example SN 2016jhr \citep{Jiang2017a} and SN 2018byg \citep{de2019a}.
These objects showed early peaks in their light curves, suggested to be
the result of radioactive material produced in the outer layers of the ejecta.
This has been predicted as a signature of a double detonation by \citet{Noebauer2017a}.
Here, however, we consider a different potential signature of a double detonation.
We investigate whether He spectral features could be directly detected.
Double detonation simulations predict 
fairly large amounts of unburnt helium at high velocities
in the outer layers of the SN ejecta \citep{fink2010a, shen2014b, gronow2020a, gronow2021a}.
In each of the double detonation models presented by \citet{fink2010a}, approximately
half of the initial helium shell mass was ejected as unburnt helium.
Radiative transfer simulations have not been able to predict whether helium features
should be able to form for the recent generation of double detonation simulations considering low
mass helium shells \citep[e.g.][]{kromer2010a, woosley2011b, townsley2019a, polin2019a, gronow2020a, shen2021b, collins2022b, pakmor2022a},
as these studies did not carry out full non-local thermodynamic equilibrium (non-LTE) simulations including non-thermal processes.

The ionisation and excitation rates of helium are strongly affected by collisions with non-thermal
electrons,
and therefore to make predictions of helium spectral features,
non-LTE radiative transfer simulations are required, and must account for non-thermal processes.
Non-thermal electrons are produced by $\gamma$-rays
from the decays of $^{56}$Ni.
The importance of collisions with non-thermal electrons on
the formation of helium features has been demonstrated for
Type Ib/Ic supernovae \citep{chugai1987a, lucy1991a, hachinger2012b}.

Non-LTE simulations by \citet{Dessart2015a} for a helium shell
detonation model by \citet{waldman2011a} clearly showed that He
spectral features form, where radiative transfer calculations are non-LTE
and include non-thermal effects.
This model, however, did not consider a secondary core detonation, and therefore the composition
of the ejecta is different to the composition predicted for the double
detonation scenario.
In particular, this model predicted large amounts of helium at all velocities, whereas
the double detonation predicts helium predominantly at high velocities \cite[see e.g.][]{fink2010a}.
The predicted spectral properties of the helium detonation model
were classified as a hybrid Type Ia/Ib due to the clear detection of helium features.

\citet{boyle2017a} investigated whether helium spectral features could form for the recent generation of
double detonation models with relatively low mass helium shells \citep[e.g.\ the models of][]{fink2010a}.
In this study, \citet{boyle2017a} used an analytic approximation to estimate the excited \ion{He}{I}
level populations,
and found that the \ion{He}{I} 10830 \AA \ and 2~$\mu$m lines may be produced by high velocity
He ($\sim$19 000 km s$^{-1}$).
They note that since the \ion{He}{I} 10830 \AA \ is substantially blueshifted,
it may be blended with the blue wing of the \ion{Mg}{II} 10927 \AA \ feature.
Further, more detailed simulations are required to confirm this result,
but this suggests that \ion{He}{I} spectral features may be observable for this scenario, and indeed recently \citet{dong2022a} showed that SN~2016dsg had an absorption line around 9700-10500 \AA \ which they find is consistent with the \ion{He}{I} 10830 \AA \ feature predicted by \citet{boyle2017a}.

\citet{liu2023a} have pointed out that SN~2022jgb and other candidate thick helium-shell detonation SN~2018byg-like SNe \citep{de2019a} show a prominent absorption feature at $\sim 1 \mu$m, which \citet{liu2023a} suggest could be high-velocity ($\sim 26000$ km s$^{-1}$) He $\lambda 10830$.
Previous works have also attempted to identify helium spectral features in the
near-IR spectra of SNe Ia \citep{meikle1996a, hoeflich1996a, mazzali1998b, marion2003a, marion2009a}.
\citet{marion2009a} present a large sample of near-IR spectra of normal SNe Ia
and do not find any clear detections of the strongest lines expected for \ion{He}{I},
at 10830 \AA \ and 2~$\mu$m.
However, they also do not detect \ion{C}{I} in any of the spectra, which has since
possibly been identified in near-IR spectra \citep{hsiao2013a, Hsiao2015a, marion2015b, wyatt2021a}.
\citet{wyatt2021a} consider whether their \ion{C}{I} detections could be misidentified \ion{He}{I}.
Following discussion of simulations by \citet{boyle2017a}, \citet{wyatt2021a} conclude that
it is unlikely the feature is \ion{He}{I}, given that in the simulations of \citet{boyle2017a},
helium absorption is visible at maximum light and grows stronger with time, whereas the
observed feature becomes weaker toward maximum light.
Additionally, the observed spectra show optical \ion{C}{II} 6580 \AA, confirming the
presence of carbon in the SN ejecta.
While the lower mass model considered by \citet{boyle2017a} does clearly show
a trend of increasing \ion{He}{I} 10830 \AA \ strength, 
the higher mass model predicts that the \ion{He}{I} 10830 \AA \ feature
remains a similar strength to the \ion{Mg}{II} 10927 \AA \ feature.
This indicates that the evolution of potential He features is likely model
dependent, and requires further study.
The approximate non-LTE treatment used by \citet{boyle2017a} assumes that
the \ion{He}{II} population is dominant.
One limiting factor for this approximation is whether the non-thermal ionisation
rate remains sufficiently high that the helium remains ionised.
Therefore, the strengths of the predicted features provide upper limits on
the potential strength of helium features.
In this work, the ionisation state is calculated self-consistently, and can therefore
better constrain the predicted strength of potential helium features.

In this paper, we investigate the potential for He spectral features to form in the double detonation explosion scenario, and whether these features could be detected in observations.
The presence of He in the ejecta is a fundamental prediction of the double detonation scenario, but it has not yet been fully addressed whether the He would produce an observable spectral feature for contemporary double detonation explosion simulations.
If an observable He spectral feature is produced, then this could be direct observational evidence for the double detonation explosion scenario, allowing us to identify the class of explosion as well as the progenitor system.
We consider an ejecta profile from a hydrodynamical simulation of the double detonation, which we describe in Section~\ref{sec:model}.
In Section~\ref{sec:spectra}, we present the simulated spectra, focusing specifically on predictions of He spectral features.
We discuss how this simulation compares to observations of a SN Ia that showed similar behaviour to our simulation in Section~\ref{sec:iPTF13ebh}, suggesting that He may have already been detected in SNe Ia, but misidentified.

\section{Methods}

\subsection{Non-LTE and non-thermal radiative transfer}

We use the time-dependent multi-dimensional Monte Carlo radiative transfer
code \textsc{artis} (\citealt{sim2007b, kromer2009a, bulla2015a}, based on the
methods of \citealt{lucy2002a, lucy2003a, lucy2005a}).
\textsc{Artis} has been extended by \citet{Shingles2020a} to solve the non-LTE
population and ionisation balance, including ionising collisions with fast non-thermal electrons.
\textsc{artis} solves the Spencer-Fano equation (as framed by \citealt{kozma1992a}) to trace the energy
distribution of high-energy leptons seeded by high-energy particles emitted in radioactive decays.
Similarly to \citet{Shingles2020a}, we allow Auger electrons
to contribute to heating, ionisation, and excitation.
We do not include excitation of
bound electrons by non-thermal collisions, since this is particularly computationally demanding, however, \citet{hachinger2012b} show that their results for He are rather insensitive to the excitation rates.

For photoionisation, \textsc{artis} uses the full Monte Carlo photon-packet
trajectories (following \citealt{lucy2003a})
to obtain a rate estimator for each level pair. For all other atomic processes,
a parameterised radiation field is used to estimate transition rates.
We use an atomic dataset based on the compilation of \textsc{cmfgen}
\citep[][see \citealt{Shingles2020a} for details]{Hillier1990a, Hillier1998a}.

The radiative transfer simulations are carried out between 2 and 32 days after explosion,
using 111 logarithmically spaced time steps.
The initial time steps are in LTE and at time step 12
(2.7 days) the non-LTE solver is switched on.

\subsection{Double detonation ejecta model}
\label{sec:model}
\begin{figure*}

\includegraphics[width=0.8\textwidth]{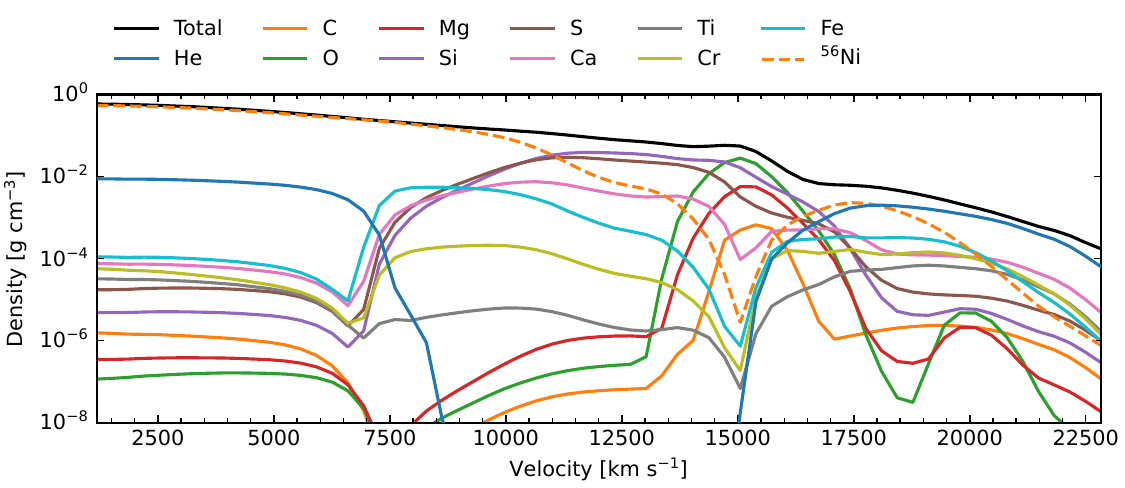}

\caption{Model densities at 100 seconds after explosion.}

\label{fig:initcomposition}
\end{figure*}

The {ejecta} model we consider in this work is based on the 3D double detonation
explosion model M2a, described by \citet{gronow2020a}.
{The light curves predicted for this model were of normal SNe Ia brightness, similar to SN~2011fe \citep{nugent2011a}.
We therefore select this model to investigate the potential for He spectral features to form in normal brightness SNe Ia.}
We construct a 1D model from a slice of model M2a in an equatorial line of sight.
Although \textsc{artis} can perform multi-dimensional simulations,
the performance cost is substantial and requires sacrifices to the number of detailed photoionisation estimators.
We find that at early times, while the ejecta are optically thick, this 1D model produces light curves and spectra similar
to those expected from a 3D simulation in this line of sight.
However, after maximum light, as the ejecta become more optically thin,
we find that this model is no longer such a good representation of the 3D line of sight.
Therefore, in this work we limit our calculations to the first few weeks after explosion.
Since the He is predominantly found in the outer ejecta, we expect that
He features are more likely to form at early times.
We show the composition of this model in Figure \ref{fig:initcomposition}.
We note that low density, outer ejecta ($v > 23200$ km s$^{-1}$) are ignored since these quickly become optically thin and are not expected to affect the spectrum at the times we consider here.
This allows for better spatial resolution of the ejecta at velocities important in forming the spectra.
The ejecta in the equatorial line of sight of Model M2a that we have based our ejecta on extend to $\sim 25000$ km s$^{-1}$, however, in the polar lines of sight the ejecta extend to higher velocities of $> 30000$~km~s$^{-1}$ (see \citealt{gronow2020a}).

The initial WD mass of Model M2a was 1.05 M$_\odot$, with a He shell
of 0.073 M$_\odot$ (see \citealt{gronow2020a} for model details).
After explosion, the total mass of He in the ejecta of Model M2a was 0.028~M$_\odot$, and the mass of $^{56}$Ni was 0.58 M$_\odot$.
In the 1D model we consider here the He mass is 0.018 M$_\odot$, and the $^{56}$Ni mass is 0.49~M$_\odot$.

The radiative transfer simulations carried out by \citet{gronow2020a}
did not include non-thermal ionisation, and made approximate non-LTE calculations.
Therefore they could not address whether helium spectral features would be expected
to form for these models.
In this work we carry out a full non-LTE simulation and include non-thermal
ionisation (as implemented by \citealt{Shingles2020a}), with the aim of investigating the effect of the more detailed 
treatment of ionisation on the formation of He features for the double detonation
scenario.
In this work, we investigate only one model due to the high computational expense of these
simulations.

The double detonation scenario leads to strong asymmetries, and therefore, in future, 3D simulations should be carried out to investigate observer angle dependencies.
Additionally, the ejecta structure and synthesised abundances show strong variations with different masses of the initial He shell and C/O core \citep[e.g.][]{fink2010a, gronow2021a, boos2021a}.
It should also be investigated how the predicted He spectral features would vary in response to such differences in the ejecta structure.

\begin{figure}
    \includegraphics[width=0.5\textwidth]{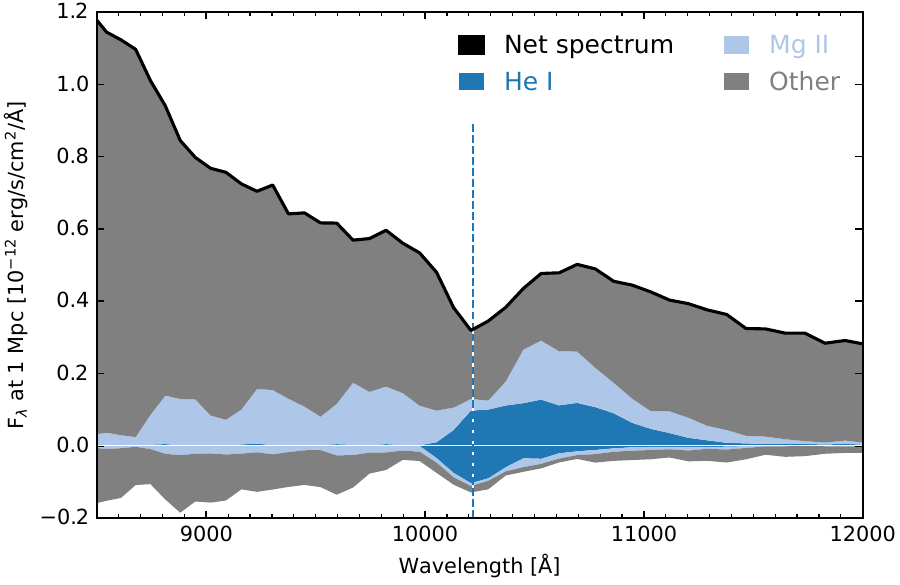}

    \caption{Spectrum at 5 days after explosion calculated using \textsc{artis} including a
    non-thermal treatment.
    We show the contributions of \ion{He}{I} and \ion{Mg}{II} to the spectrum, and indicate the absorption by these ions beneath the axis.
    The wavelength of the minimum of the \ion{He}{I}~10830~\AA \ feature is indicated by the dashed line. 
    We note that no \ion{C}{I} feature is formed.
    }
    \label{fig:specemission5days}
\end{figure}

\begin{figure}
    \centering
    \begin{subfigure}[b]{0.48\textwidth}
        \includegraphics[width=1\textwidth]{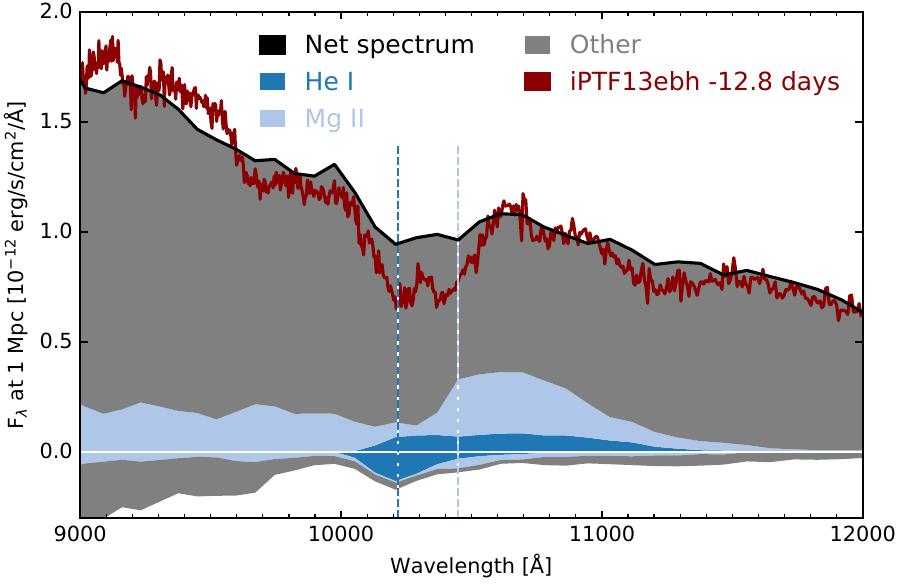}
        \caption{8 days after explosion.}
        \label{fig:specemission8days}
    \end{subfigure}
    \smallskip

    \begin{subfigure}[b]{0.48\textwidth}
        \includegraphics[width=1\textwidth]{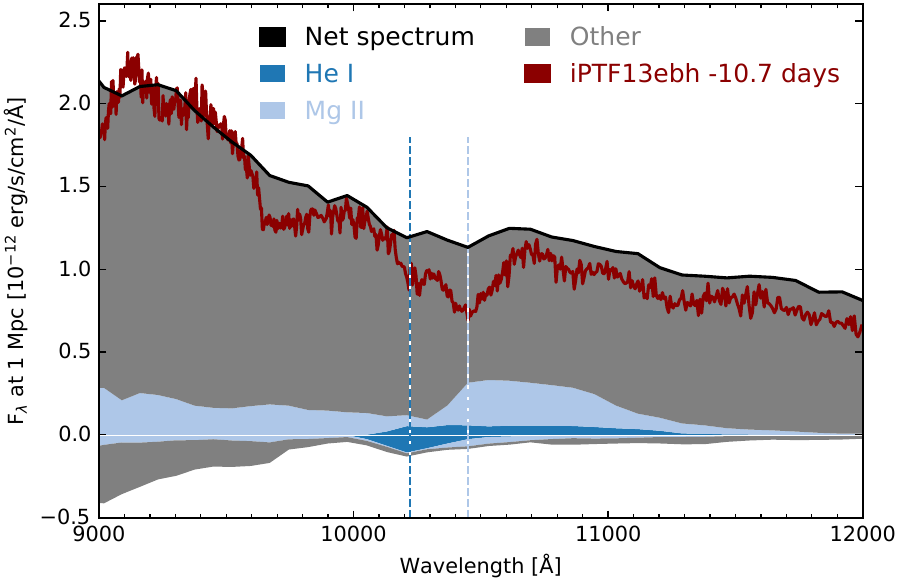}
        \caption{10 days after explosion.}
        \smallskip
        \label{fig:specemission11days}
    \end{subfigure}

    \begin{subfigure}[b]{0.48\textwidth}
        \includegraphics[width=1\textwidth]{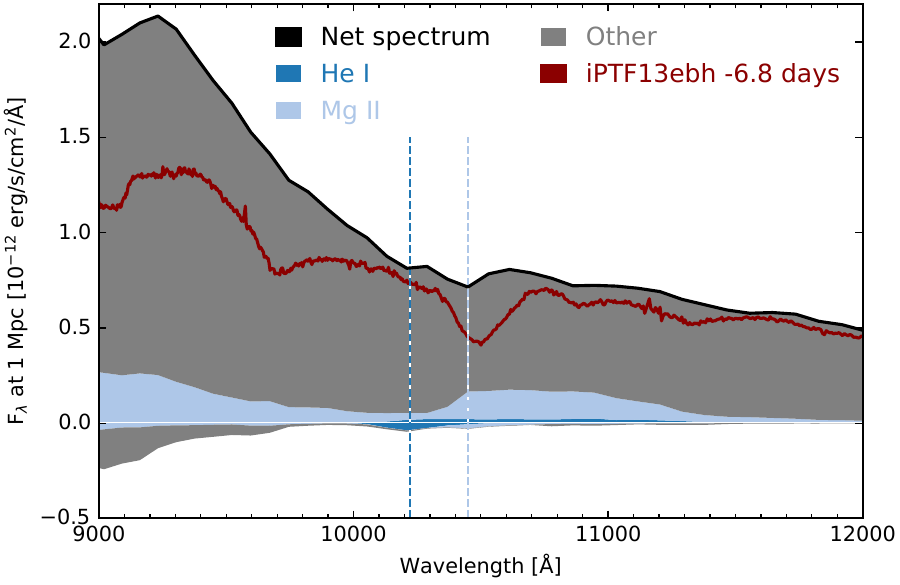}
        \caption{14 days after explosion.}
        \label{fig:specemission14days}
    \end{subfigure}

    \caption[]{
    Model spectra sequence showing the evolution of the \ion{He}{I} 10830~\AA \ feature with time. 
    The simulated time since explosion is noted below each model spectrum.
    The wavelengths of the minima of the \ion{He}{I} 10830 \AA \ and \ion{Mg}{II} 10927 \AA \ features are indicated by the dashed lines.
    We compare our model spectra to iPTF13ebh \citep{Hsiao2015a}.
    Since the time of explosion for this object is uncertain we list the observed times relative to B-band maximum of iPTF13ebh. 
    The observed spectra have been scaled to be of similar brightness to our simulated spectra.
    We note that our model does not predict a \ion{C}{I} feature.
    }

\end{figure}

\section{Results}

\subsection{Helium spectral features}
\label{sec:spectra}

In future work we will present the full results of this non-LTE simulation,
but in this paper we limit discussion to the ability of this simulation to
produce He spectral features.
We note that a level of Monte Carlo noise is present in our model spectra.

{We identify the presence of He features in our simulated spectra by tagging escaping Monte Carlo radiation packets with the ion with which they last interacted.
By examining packets whose last interaction was with He, we can quantify the contribution of He to the synthetic spectra and identify the specific spectral features for which He is responsible.}

\begin{figure*}
    \centering
    \begin{subfigure}[b]{0.6\textwidth}
        \includegraphics[width=1\textwidth]{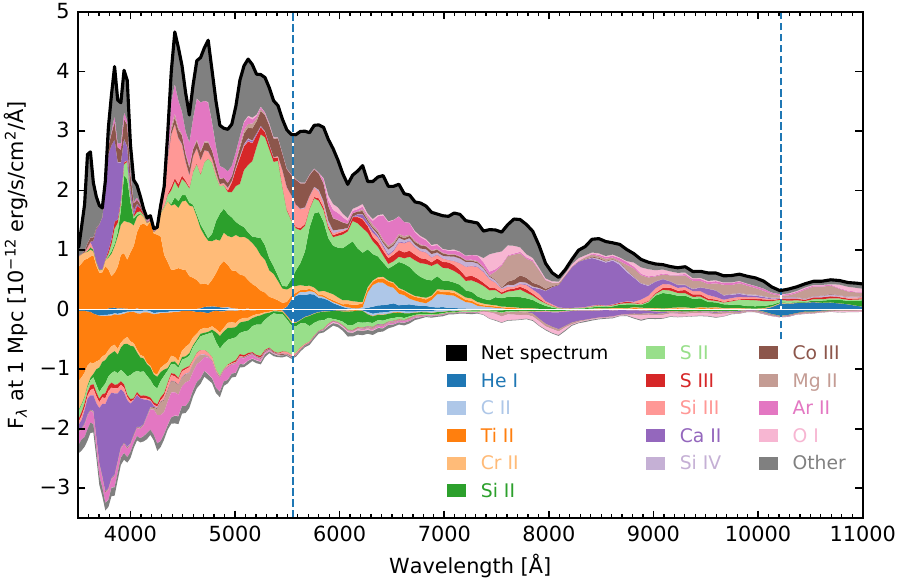}
        \caption{5 days after explosion. The dashed lines indicate the wavelengths of the \ion{He}{I} absorption.}
        \label{fig:specemission5daysoptical}
    \end{subfigure}

    \begin{subfigure}[b]{0.6\textwidth}
        \includegraphics[width=1\textwidth]{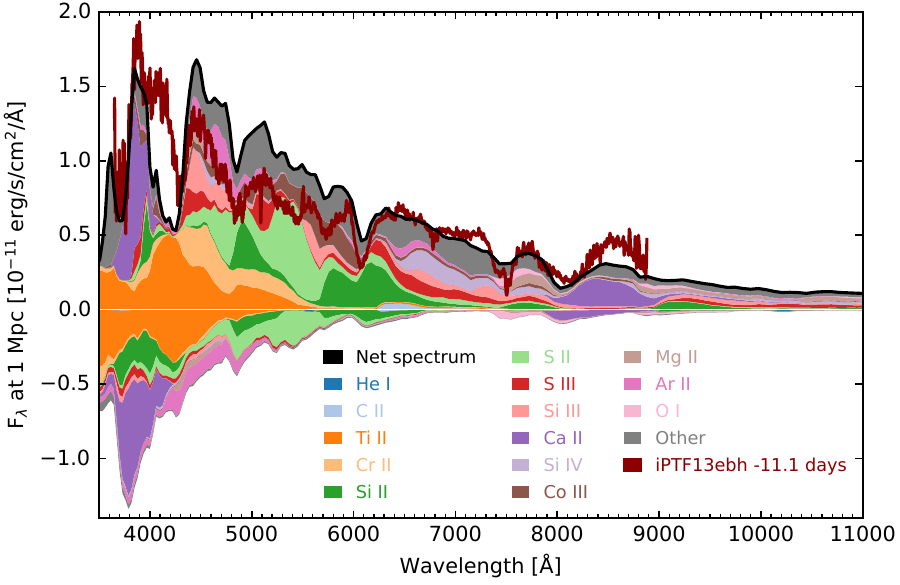}
        \caption{9 days after explosion.}
        \label{fig:specemission8daysoptical}
    \end{subfigure}

    \caption[]{Model optical spectra where the contributions of key species are indicated.
    We compare the spectrum at 9 days after explosion to the spectrum of iPTF13ebh (at 11.1 days before the observed B-band maximum light).
    The observed spectrum has been scaled to be of similar brightness to the simulated spectrum.
    }

\end{figure*}

\subsubsection{NIR \ion{He}{I} $\lambda$10830}

We predominantly focus on the wavelength region around where the \ion{He}{I} 10830 \AA \ (2$^3$S - 2$^3$P) feature would form.
We expect this to be the strongest He feature with the least blending from other spectral lines.
In Figure \ref{fig:specemission5days} we show the spectrum at these wavelengths at 5 days after explosion, and indicate the contributions of \ion{He}{I} to the spectrum.
In the radiative transfer simulation, we record details of the last interaction each
Monte Carlo packet underwent before escaping the ejecta. For each wavelength bin in the
synthetic spectrum the area under the spectrum is colour coded in proportion to the energy
carried by packets in that wavelength bin whose last interaction was with each of the ions
considered.
We also construct an equivalent histogram based on
where the wavelength bin packets were prior to their last interaction (i.e. indicating where
packets last underwent absorption/scattering/fluorescence) and plot this on the negative axis
under the spectrum to indicate the key absorption processes.
We find that the \ion{He}{I} 10830~\AA \
feature is expected to form for this model, and that a relatively strong He absorption feature forms in the first few days 
after explosion.
Over time the strength of this feature weakens, and at times later than $\sim$2
weeks after explosion this feature is no longer visible 
(see the spectral time series shown by
Figures \ref{fig:specemission8days}, \ref{fig:specemission11days}, and \ref{fig:specemission14days}).

In Figure \ref{fig:specemission5days}, showing the spectrum at 5 days after explosion,
the \ion{He}{I} 10830 \AA \ feature is initially completely blended with the
\ion{Mg}{II} 10927~\AA \ feature.
In observations, this could be mistaken for a broad \ion{Mg}{II} feature.
In the region around the 10830 \AA \ feature, the emission from other lines
is relatively weak,
hence we find the 10830~\AA \ feature
to be relatively strong.
The `Other' elements in Figure \ref{fig:specemission5days} are predominantly intermediate mass elements, with the strongest contributions being from Si and S.

By 8 days after explosion, the \ion{He}{I} feature has separated from the \ion{Mg}{II} feature
and is clearly distinguishable at the blue wing
of the \ion{Mg}{II} feature.
The wavelength bin sizes have been chosen to minimise Monte Carlo noise in the model spectra. We note that the \ion{He}{I} feature separating from the \ion{Mg}{II} feature can be seen when using smaller wavelength bins.
At 10 days after explosion, a distinct secondary feature to the blue wing of the
\ion{Mg}{II} feature can also be seen in Figure~\ref{fig:specemission11days},
although this is weaker than at 8 days after explosion.
At 5 days after explosion, the \ion{He}{I} 10830 \AA \ feature
is formed at $\sim$19500~km/s,
while the \ion{Mg}{II} 10927 \AA \ feature is formed at $\sim$16000~km/s.
Over time, the velocity of the \ion{He}{I} 10830 \AA \ feature decreases.
At 2 weeks after explosion the feature is formed at $\sim$18000~km/s.

Such features to the blue wing of the \ion{Mg}{II} feature have previously been
detected \citep{hoeflich2002b, hsiao2013a, Hsiao2015a, marion2015b},
but have been identified as \ion{C}{I} 10693 \AA.
This will be discussed further in Section~\ref{sec:NIR-CI}.
By 2 weeks after explosion, a clear \ion{He}{I} 10830 \AA \ feature can no longer be seen, however, the blue wing of the 
\ion{Mg}{II} feature is flattened due to \ion{He}{I} 10830 \AA.

This demonstrates that NIR \ion{He}{I} 10830 \AA \ could be a signature of the double detonation explosion scenario, and that early NIR spectra are key to detecting the presence of this feature.

\subsubsection{Optical He}

We now discuss the presence of He features in the model optical spectra.
At 5 days we find the \ion{He}{I} 5876 \AA \ feature to be the strongest optical He feature (see Figure~\ref{fig:specemission5daysoptical}),
although, it is blended with other strong lines,
predominantly from \ion{S}{II} and \ion{Si}{II}.
A weak \ion{He}{I} absorption feature is found at $\sim 5555$ \AA, however, this is blended with \ion{S}{II} absorption.
Similarly to the 10830~\AA \ feature, this is also strongest at the earliest times, and
becomes weaker over time, however the 5876~\AA \ feature fades more rapidly than
the 10830 \AA, and is no longer formed by $\sim$ 9 days after explosion.
Figure~\ref{fig:specemission8daysoptical} shows the key species contributing
to the optical spectrum at 9 days after explosion, where contributions
from \ion{He}{I} can no longer be seen around 5876~\AA.
This shows that the \ion{He}{I} 10830 \AA \ feature can be relatively strong
without the expectation that the 5876~\AA \ feature should also be strong.
This was also noted by \citet{mazzali1998b}, and
\citet{nugent1997a} found no strong evidence for optical He lines in their calculations
for a helium detonation model.
We note that we find no \ion{He}{II} features in our simulation.

\subsubsection{\ion{He}{I} 2 $\mu$m}

Our model spectra show a high level of Monte Carlo noise in the wavelength region around 2 $\mu$m, where we would expect the \ion{He}{I} 2 $\mu$m feature to form.
{Increasing the number of Monte Carlo packets in our simulation to improve the signal to noise at such red wavelengths is prohibitively computationally expensive at this time.}
We therefore cannot predict whether this model would produce a clear \ion{He}{I} 2 $\mu$m feature.
However, qualitatively, we find that there are contributions to the spectrum from \ion{He}{I} in this wavelength region at very early times which decrease in strength with time and have faded by $\sim 1$ week after explosion.
We therefore suggest that it is possible that a \ion{He}{I} 2~$\mu$m feature could form within the first week after explosion, however we cannot confirm this with our current simulation.

\subsection{Discussion}

We have shown for a particular instance of a double detonation model that the formation of a clear \ion{He}{I} 10830 \AA \ feature is predicted by our radiative transfer calculations.
Future simulations should investigate whether this is a prediction of all double detonation models.
Importantly, we do not expect a clear He feature in the optical spectra, implying that the non-detection of an optical He spectral feature can not be used to rule out the presence of He when identifying a \ion{He}{I} 10830 \AA \ feature.

The He shell in this model before ignition was relatively massive (compared to that considered by 
\citealt{townsley2019a}, or the minimal He shell models presented by \citealt{gronow2021a} and \citealt{boos2021a}),
and so has more He in the ejecta
at a more extended range of velocities than would be synthesised for a model with a
thinner He shell.
{Additionally, the 3D model M2a \citep{gronow2020a} on which our 1D model is based showed strong asymmetries.
For example, He extends to higher velocities along the negative polar axis (see \citealt{gronow2020a}, figure 13), in which case a He feature may form at higher velocities.}
Investigating the dependence of the strength {and velocity} of He features on the choice of model {and observer orientation}
is beyond the scope of this work,
however, it is likely that both the mass of
He, and the mass of $^{56}$Ni mixed with the He will influence the
predicted He spectral features.
{In our model, $^{56}$Ni is synthesised in the He shell detonation, and the presence of $^{56}$Ni in the ejecta in the same regions as the He likely increases the number of non-thermal collisions with He.
Future simulations should investigate the sensitivity of the line formation to the degree of mixing of $^{56}$Ni with the He.}

It is interesting that we find the He features become weaker with time, given
that in simulations of SNe Ib this feature is found to grow in strength with time
\citep{hachinger2012b, teffs2020a}, and also grows in strength in the
He shell detonation model of \citet{Dessart2015a}.
The He in the SNe~Ib simulations, and in the He shell detonation model,
extends over a wide range of
velocities.
In our simulation, however, the He is limited to a smaller range of velocities (see Figure~\ref{fig:initcomposition}) in the outer ejecta.
In the model considered by \citet{Dessart2015a} He is the
most abundant element at all velocities (see their figure 1) and they have more
He by mass than in our model.
\citet{Dessart2015a} find that \ion{He}{I} 10830 \AA \ is strong
at all times beyond 5 days.
They also find \ion{He}{I}~5875~\AA \ is clearly visible
between 5--15 days until line blanketing by \ion{Ti}{II} causes the feature to
fade.
It is possible that the \ion{He}{I} 10830 \AA \ feature remained visible at all times
due to the high abundance of He present at all velocities, and that the strengths
of the \ion{He}{I} features were found to be so strong due to the higher abundance
of He in the model.
Future work should investigate how model structure, and masses of He affect predictions
of He spectral features.

\citet{boyle2017a} consider both a low-mass model
(0.58 \msun \ CO core and 0.21 \msun \ He shell)
and a high-mass model (1.025 \msun \ CO core and 0.055 \msun \ He shell).
Like our model, the high-mass model shows a deficit of He at intermediate velocities,
however, their low-mass model does not show this
(see figure 3 of \citealt{boyle2017a}).
Their low-mass model predicts that the \ion{He}{I} 10830 \AA \ feature
clearly grows in strength, and becomes much stronger than the \ion{Mg}{II} 10927~\AA \ feature.
In the high-mass model, however, the \ion{He}{I} 10830 \AA \ feature becomes
slightly stronger with time,
but remains similar in strength to the \ion{Mg}{II} 10927 \AA \ feature
(figure 11 of \citealt{boyle2017a}).
The approximate non-LTE treatment used by \citet{boyle2017a} assumed that the \ion{He}{II} population is dominant.
In our simulation we find that indeed \ion{He}{II} is the dominant ionisation state in the outer ejecta throughout our simulation.
Generally, \ion{He}{I} is the next most highly populated in the outer ejecta.

\subsubsection{NIR \ion{C}{I}}
\label{sec:NIR-CI}

A number of detections of a feature to the blue wing of the \ion{Mg}{II}
feature {(similar to our simulated \ion{He}{I} 10830 \AA \ feature)} have been reported, which have previously been suggested to be
\ion{C}{I}.
Specifically, clear features were identified in SN 1999by \citep{hoeflich2002b}, iPTF13ebh \citep{Hsiao2015a}, SN~2015bp \citep{wyatt2021a} and
SN~2012ij \citep{li2022a},
and an additional feature affecting the blue wing of the \ion{Mg}{II}
feature has been reported in
SN~2011fe \citep{hsiao2013a} and SN~2014J \citep{marion2015b}.

\citet{Hsiao2015a} claim the feature in the blue wing 
of the \ion{Mg}{II} feature in iPTF13ebh is \ion{C}{I}
(see Section~\ref{sec:iPTF13ebh} below).
\citet{wyatt2021a} and \citet{liu2023a} have detected similar features in SN~2015bp and SN~2012ij respectively, for which they also propose identification with 
\ion{C}{I}.
These objects were classified as transitional SNe Ia, showing similar spectral properties to 91bg-like SNe, but also showing secondary NIR maxima, 
which also includes objects such as
SN~1986G \citep{phillips1987a}, SN~2003gs \citep{krisciunas2009a} and SN~2004eo \citep{pastorello2007b, mazzali2008b}.
These have been suggested to link 91bg-like SNe to normal SNe Ia.
\citet{hoeflich2002b} also identify the feature in SN~1999by as \ion{C}{I}.
SN~1999by is a 91bg-like, subluminous object, and
the spectral evolution of this feature is similar to that in iPTF13ebh.

\ion{C}{I} has tentatively been identified in the NIR spectra of the normal
SN~2011fe \citep{hsiao2013a}, which showed a flattened wing to the 
\ion{Mg}{II} feature.
SN~2014J also showed a flattened wing similar to that in SN~2011fe, and 
was shown by \citet{marion2015b} to be a possible detection of \ion{C}{I}.
The feature in SN~2011fe showed an increase in strength towards maximum, 
however in iPTF13ebh it rapidly decreased in strength.

In our simulation we do not find any \ion{C}{I} features.
We do find contributions to the optical spectra from \ion{C}{II} 6580 \AA \ and 7235~\AA,
but these are weak and would not be easily detectable as they are heavily blended with
stronger lines (see Figure~\ref{fig:specemission5daysoptical}).
At the times when our \ion{He}{I} feature forms, even in the outer layers of our simulation ejecta where temperatures are coolest and the ejecta are the least highly ionised, \ion{C}{II}
is the dominant ionisation stage, with \ion{C}{III} being the next most abundant, indicating that in this region \ion{C}{III} is recombining to \ion{C}{II}. 
\ion{C}{I} is significantly depopulated relative to both \ion{C}{II} and \ion{C}{III} (on the order of 5 magnitudes lower ionisation fraction).
As such, we would not expect \ion{C}{I} features to form.
This is consistent with e.g.\ \citet{tanaka2008a} and \citet{heringer2019a} who also found
\ion{C}{II} to be the dominant ionisation state, rather than \ion{C}{I}.
However, note that some simulations have found that \ion{C}{I} features can form \citep[e.g.][]{hoeflich2002b, Hsiao2015a}.

\subsubsection{{Comparison of simulated spectra to observations}}
\label{sec:iPTF13ebh}

\begin{table}
    \caption{Peak light curve brightnesses for iPTF13ebh from \citet{Hsiao2015a},
    compared to the values calculated for our model.}
  \label{tab:iPTF13ebhpeakbrightnesses}
  \centering
  \begin{tabular}{@{}lll}
  \hline
                                     & iPTF13ebh                          & Model \\ \hline
M$_{\mathrm{u,max}}$ & -18.34 ± 0.19 & -18.2                        \\
M$_{\mathrm{B,max}}$ & -18.95 ± 0.19 & -18.9                        \\
M$_{\mathrm{V,max}}$ & -19.01 ± 0.18 & -19.1                        \\
M$_{\mathrm{g,max}}$ & -19.03 ± 0.18 & -19.1                        \\
M$_{\mathrm{r,max}}$ & -18.99 ± 0.18 & -18.8                        \\
M$_{\mathrm{i,max}}$ & -18.52 ± 0.18 & -18.3                        \\
M$_{\mathrm{J,max}}$ & -18.75 ± 0.18 & -18.4                        \\
M$_{\mathrm{H,max}}$ & -18.59 ± 0.18 & -18.7 \\ \hline
\end{tabular}
\end{table}

We compare our {simulated} spectra to those of iPTF13ebh \citep{Hsiao2015a}, which showed a clear
second component to the blue wing of the \ion{Mg}{II} feature in the early 
NIR spectra, and was identified as \ion{C}{I} 10693~\AA.
We chose this object given the high quality NIR spectra obtained at early times.
It also showed similar peak brightnesses to our model
(see Table~\ref{tab:iPTF13ebhpeakbrightnesses}).
{We note that we have not tuned our model in any way to try to match the spectra of iPTF13ebh. We are comparing our prediction based on an existing double detonation ejecta model to 
the observations of iPTF13ebh given the similarities between our simulated \ion{He}{I} 10830 \AA \ feature and the observed feature (previously identified as \ion{C}{I}) in iPTF13ebh.}

iPTF13ebh has a $\Delta$\textit{m}$_{15}$(B) value of 1.79 and is therefore in the
fast-declining SN Ia category, but is classified as a transitional event:
the photometric properties of iPTF13ebh place it in the category of normal SNe Ia,
however its spectral properties resemble those of SN 1999by which is a member of
the 91bg-like class.
iPTF13ebh however, showed no apparent \ion{Ti}{II} features
and is therefore not considered a 91bg-like object.

\citet{Hsiao2015a} found no optical \ion{C}{I} features in iPTF13ebh.
They did, however, find evidence for a weak optical
\ion{C}{II} 6580 \AA \ feature at approximately the same epoch,
which they also found to weaken rapidly.
They propose identification of \ion{C}{I} 9093 \AA \ as weak notches in the two earliest
spectra, and 
suggest that \ion{C}{I} 9406 \AA \ and 11754 \AA \ features could be present
in the spectra, but blending makes these detections uncertain.

Since the observed feature at the blue wing
of the \ion{Mg}{II} 10927~\AA \ feature in iPTF13ebh shows similar behaviour to the simulated high velocity \ion{He}{I} 
10830~\AA \ feature in our model and the feature, we compare our model 
to the early NIR spectra of iPTF13ebh
\citep{Hsiao2015b}.
All spectra have been dereddened and redshift corrected.

We plot the spectra of iPTF13ebh
in Figures \ref{fig:specemission8days}, 
\ref{fig:specemission11days} and \ref{fig:specemission14days},
along with our model spectra.
We use times relative to maximum light for the observed spectra since the inferred time since explosion is uncertain (see \citealt{Hsiao2015a}).
In the spectra of iPTF13ebh the feature can be seen decreasing in strength
over time until in Figure~\ref{fig:specemission14days} at 6.8 days before maximum it is barely visible.
Maximum light in B-band occurs in our simulation at $\sim$18.6 days after explosion.
The \ion{He}{I} feature predicted in our simulation is able to
reproduce the blue wing of the \ion{Mg}{II} feature in all three spectra, although at $\sim$ 2.2 days earlier than the times of the observed spectra.
The velocities of the \ion{He}{I} and \ion{Mg}{II} features appear
to be well matched, as well as the trend in strength of the \ion{He}{I} feature relative to the \ion{Mg}{II} feature.

Given this match between the evolution of our \ion{He}{I} 10830 \AA \ feature and the feature in iPTF13ebh, as well as the lack of \ion{C}{I} in our model spectra,
we identify this feature as \ion{He}{I} $\lambda 10830$, and suggest that this could be a signature of the double detonation scenario.

We show a comparison between the optical spectrum of our model at 9 days after explosion to the observed spectrum of iPTF13ebh in Figure~\ref{fig:specemission8daysoptical}.
Our model shows a \ion{Ti}{II} trough at $\sim$ 4000--4500 \AA \ which is not observed in iPTF13ebh, but apart from this feature the spectra show reasonable agreement.
At this time our model does not predict any optical \ion{He}{I} features, which is consistent with the observations of iPTF13ebh.
{We note that the spectra of iPTF13ebh have strong telluric absorption in the region around where the \ion{He}{I} 2 $\mu$m feature could form \citep{Hsiao2015a}, and therefore we can not identify whether this feature is present in the observed spectra.}

\section{Conclusions}

{We have shown for an instance of a double detonation explosion model that \ion{He}{I} 10830 \AA \ is a predicted spectral feature.}
By including a non-thermal solution in our
calculations, \ion{He}{I} features are able to form for the double detonation
scenario, and could be a signature of this explosion mechanism.

In our simulation, we find a clear absorption feature due to high velocity
\ion{He}{I} 10830 \AA.
We find the \ion{He}{I} 10830 \AA \ feature is strongest in the first few days after explosion, and becomes
weaker over time.
After 2 weeks, this feature is no longer visible.
The \ion{He}{I} 10830 \AA \ feature is initially blended 
with \ion{Mg}{II}, and therefore could potentially be concealed in observed spectra.
We find that over time the \ion{He}{I} feature separates from the \ion{Mg}{II},
such that we see two distinct features.
 We note, however, it is likely that the strength and evolution of the \ion{He}{I} $\lambda 10830$ feature is model dependent, and future studies should be carried out investigating different mass models.

We find that the \ion{He}{I} 5876 \AA \ feature is able to form in our simulation at very early times, however, this is
blended with other strong lines and quickly fades.
Therefore, the non-detection of this feature in the optical spectra should not be used to rule out the presence of the \ion{He}{I} 10830 \AA \ feature, particularly later than the first week after explosion.

Due to Monte Carlo noise in our simulation around 2 $\mu$m we cannot confirm whether a clear \ion{He}{I} 2 $\mu$m feature would form for this model.
However, we do find \ion{He}{I} contributions to the model spectra in this wavelength range indicating that a \ion{He}{I} 2 $\mu$m feature could potentially form that would be strongest at early times and decrease in strength until $\sim 1$ week after explosion.

{In addition to our result that \ion{He}{I} 10830 \AA \ is a predicted spectral feature for the double detonation scenario, we compare the simulated evolution of this feature to observations of iPTF13ebh, which we suggest showed evidence of a \ion{He}{I} 10830 \AA \ feature.}
A secondary component at the blue wing of the \ion{Mg}{II} $\lambda 10927$ feature has previously been observed in transitional SNe Ia, such as iPTF13ebh \citep{Hsiao2015a}, which showed a similar evolution to that predicted by the \ion{He}{I} 10830 \AA \ feature in our model.
However, this has previously been identified as \ion{C}{I} 10693~\AA.
We have compared our model spectra to the observations of iPTF13ebh and found that \ion{He}{I} 10830 \AA \ can reproduce this feature,
and we suggest that this is evidence for \ion{He}{I} in SNe Ia and for the double detonation explosion scenario.
This shows that \ion{He}{I} may not always be concealed in the 
spectra of SNe Ia.
In order to potentially detect the \ion{He}{I} 10830 \AA \ in observations,
and to identify this feature,
early NIR observations are required.

\section*{Acknowledgements}

CEC is grateful for support from the Department for the Economy (DfE) and acknowledges support by the European Research Council (ERC) under the European Union’s Horizon 2020 research and innovation program under grant agreement No. 759253.
CEC and LJS acknowledge support by Deutsche Forschungsgemeinschaft (DFG, German Research Foundation) - Project-ID 279384907 - SFB 1245 and MA 4248/3-1.
The work of SAS was supported by the Science and Technology Facilities Council [grant numbers ST/P000312/1, ST/T000198/1, ST/X00094X/1].
LJS acknowledges support by the European Research Council (ERC) under the European Union’s Horizon 2020 research and innovation program (ERC Advanced Grant KILONOVA No. 885281).
The work of SG and FKR was supported by the Deutsche Forschungsgemeinschaft (DFG, German Research Foundation) under Project-ID 138713538, SFB 881 (“The Milky Way System”, Subproject A10), by the ChETEC COST Action (CA16117), by the National Science Foundation under Grant No. OISE-1927130 (IReNA), and by the Klaus Tschira Foundation.
The authors gratefully acknowledge the Gauss Centre for Supercomputing e.V. (www.gauss-centre.eu) for funding this project by providing computing time on the GCS Supercomputer JUWELS at J{\"u}lich Supercomputing Centre (JSC).
This work was performed using the Cambridge Service for Data Driven Discovery (CSD3), part of which is operated by the University of Cambridge Research Computing on behalf of the STFC DiRAC HPC Facility (www.dirac.ac.uk). The DiRAC component of CSD3 was funded by BEIS capital funding via STFC capital grants ST/P002307/1 and ST/R002452/1 and STFC operations grant ST/R00689X/1. DiRAC is part of the National e-Infrastructure.
CEC is grateful for computational support by the VIRGO cluster at GSI.
NumPy and SciPy
\citep{oliphant2007a}, Matplotlib
\citep{hunter2007a} and \href{https://github.com/artis-mcrt/artistools}{\textsc{artistools}}\footnote{\href{https://github.com/artis-mcrt/artistools/}{https://github.com/artis-mcrt/artistools/}}\citep{shingles-artistools} were used for data processing and plotting.


\section*{Data Availability}

The spectra presented here will be made available on
the Heidelberg supernova model archive
\href{https://hesma.h-its.org}{\textsc{hesma}}\footnote{\href{https://hesma.h-its.org}{https://hesma.h-its.org}}
\citep{kromer2017a}.




\bibliographystyle{mnras}
\bibliography{astrofritz} 






\bsp	
\label{lastpage}
\end{document}